\documentclass[prl,twocolumn,superscriptaddress]{revtex4}
\usepackage{graphicx}
\usepackage{color}
\usepackage{amsmath}
\usepackage{color}

\def\ket#1{\mathinner{|{#1}\rangle}}

\def\text#1{\textrm{#1}}

\begin{document}

\title{Proposal for Exploring Macroscopic Entanglement \\
with a Single Photon and Coherent States}

\date{\today}


\author{Pavel Sekatski}
\affiliation{Group of Applied Physics, University of Geneva, CH-1211 Geneva 4, Switzerland}

\author{Nicolas Sangouard}
\affiliation{Group of Applied Physics, University of Geneva, CH-1211 Geneva 4, Switzerland}

\author{Magdalena Stobi\'nska}
\affiliation{Institute of Physics, Polish Academy of Sciences, Al. Lotnik\'ow 32/46, 02-668 Warsaw, Poland}
\affiliation{Institute for Theoretical Physics and Astrophysics, University of Gda\'nsk, Gda\'nsk, Poland}

\author{F\'{e}lix Bussi\`{e}res}
\affiliation{Group of Applied Physics, University of Geneva, CH-1211 Geneva 4, Switzerland}

\author{Mikael Afzelius}
\affiliation{Group of Applied Physics, University of Geneva, CH-1211 Geneva 4, Switzerland}

\author{Nicolas Gisin}
\affiliation{Group of Applied Physics, University of Geneva, CH-1211 Geneva 4, Switzerland}

\begin{abstract}
Entanglement between macroscopically populated states can easily be created by combining a single photon and a bright coherent state on a beam-splitter. Motivated by the simplicity of this technique, we report on a method using displacement operations in the phase space and basic photon detections to reveal such an entanglement. We show that this eminently feasible approach provides an attractive way for exploring entanglement at various scales, ranging from one to a thousand photons. This offers an instructive viewpoint to gain insight into the reasons that make it hard to observe quantum features in our macroscopic world.\end{abstract}
\maketitle

\paragraph{Introduction.}


Why do we not easily observe entanglement between macroscopically populated (macro) systems? Decoherence is widely accepted as being responsible~\cite{Zurek03}. Loss or any other form of interactions with the surroundings more and more rapidly destroys the quantum features of physical systems as their size increases. Technologically demanding experiments, involving Rydberg atoms interacting with the electromagnetic field of a high-finess cavity~\cite{Brune96} or superconducting devices cooled down to a few tens of mK~\cite{Friedman00}, have strengthened this idea.\\

Measurement precision is likely another issue~\cite{Mermin80}. In a recent experiment~\cite{DeMartini08}, a phase covariant cloner has been used to produce tens thousand clones of a single photon belonging initially to an entangled pair. In the absence of loss, this leads to a micro-macro entangled state~\cite{ampli_sti}. Nobody knows however, how the entanglement degrades with a lossy amplification~\cite{note_ampli}. This led to a lively debate~\cite{De Martini Review} concerning the presence of entanglement in the experiment reported in Ref.~\cite{DeMartini08}. What is known is that under moderate coarse grained measurements, the micro-macro entanglement resulting from a lossless amplification leads to a probability distribution of results that is very close to the one coming from a separable micro-macro state~\cite{Raeisi11}. This suggests that even if a macro system could be perfectly isolated from its environment, its quantum nature would require very precise measurements to be observed.\\

Both for practical considerations and from a conceptual point of view, it is of great interest to look for ways as simple as possible to generate and measure macro entanglement, so that the effects of decoherence processes and the requirements on the measurements can all be studied together. In this letter, we focus on an approach based on linear optics only, where a single photon and a coherent state are combined on a mere beam-splitter.
We show the resulting path-entangled state \cite{Windhager11} could be useful in quantum metrology for precision phase measurement. More importantly, it allows one to easily explore entanglement over various photon number scales, spanning from the micro to the macro domain, by simply tuning the intensity of the laser producing the input coherent state. We show that the entanglement is more and more sensitive to phase fluctuations between the paths when it grows. However, it features surprising robustness against loss, making it well suited to travel over long distances and to be stored in atomic ensembles. We further present a simple and natural method relying on local displacement operations in the phase-space and basic photon detections to reveal the entanglement. Our analysis shows that the precision of the proposed measurement is connected to the limited ability to control the phase of the local oscillator that is used to perform the phase-space displacements. We also report on preliminary experimental results demonstrating that entanglement containing more than a thousand photons could be created and measured with currently available technologies.\\

\paragraph{Creating macro entanglement by combining a single photon with a bright coherent state on a beam-splitter.} A particularly simple way of generating entanglement is to use a beam-splitter. Consider a single photon sent through a 50:50 beam-splitter. It occupies the two output modes $A$ and $B$ with the same probability and creates a simple form of entanglement between spatial modes (path entanglement)
\begin{equation}
\label{singlephotonent}
\frac{1}{\sqrt{2}}\left(|1\rangle_A|0\rangle_B-|0\rangle_A|1\rangle_B\right)
\end{equation}
known as single-photon entanglement \cite{vanEnk05}. {In fact, any product input state of the form $\rho_A \otimes |\beta\rangle\langle \beta|_B$, where $|\beta\rangle$ is a coherent state, leads to entanglement after the beam-splitter if and only if $\rho_A$ is nonclassical \cite{Asboth05, Solomon11}, i.e. it cannot be written as a mixture of coherent states \cite{Glauber63}. Hence, a mere beam-splitter links two fundamental concepts of quantum physics: non-classicality and entanglement. It also provides an attractive way for bringing entanglement to macroscopic level, as explained below.}

\begin{figure}[ht!]\includegraphics[width=9cm]{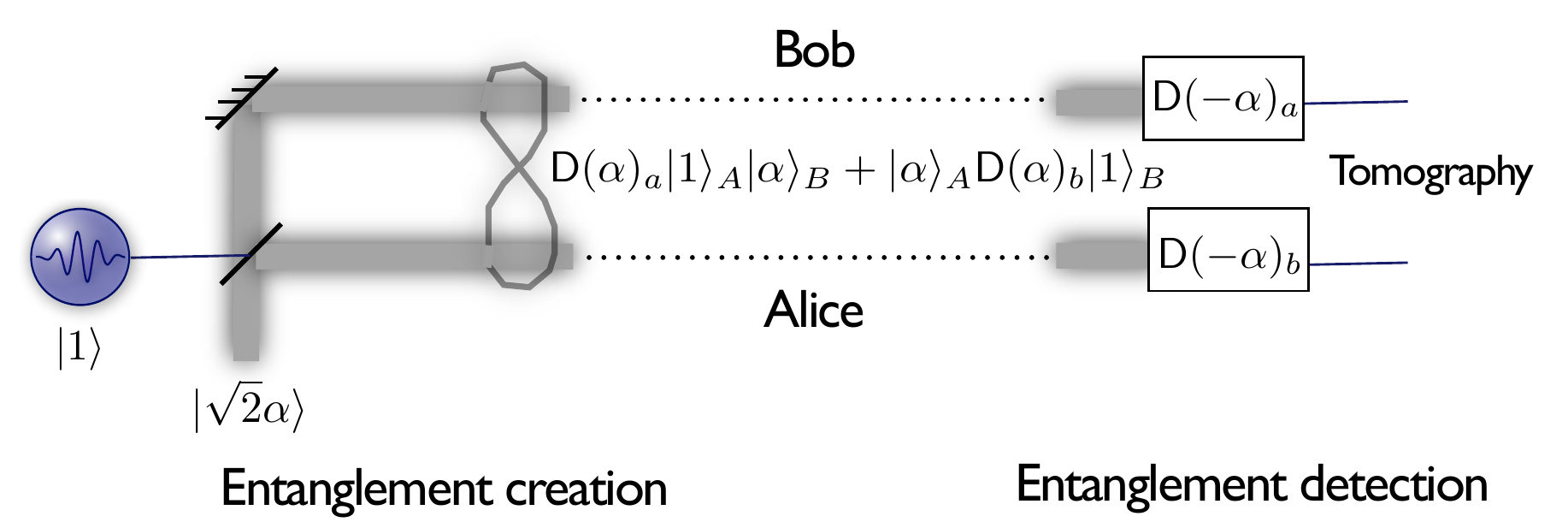}
\caption{Creation and detection of macro entanglement by combining a single photon Fock state $|1\rangle$ and a coherent state $|\sqrt{2}\alpha\rangle$ on a 50:50 beam-splitter.}
\label{fig1}
\end{figure}

Let us focus on the case where the beam-splitter inputs are a single photon $|1\rangle$ and a coherent state with $2|\alpha|^2$ photons on average (see Fig. \ref{fig1})
\begin{equation}
 |\psi_{\rm{in}}\rangle=a^\dag |0 \rangle_A\otimes \textsf{D}_b(\sqrt{2}\alpha) |0 \rangle_B.
 \end{equation}
$\textsf{D}_b(\sqrt{2}\alpha)=e^{\sqrt{2}\left(\alpha b^\dag-\alpha^\star b\right)}$ is the displacement operator generating a coherent state $|\sqrt{2}\alpha\rangle$ from vacuum (in mode B), $a$ and $b$ are bosonic annihilation operators associated with modes A and B, respectively. A 50:50 beam-splitter transforms $(a, b)$ into $((a-b)/\sqrt{2},(a+b)/\sqrt{2}).$ Since $a$ and $b$ commute, the output state reads
\begin{equation}
\label{displ_singlephent}
|\psi_{\rm{out}}\rangle= \frac{1}{\sqrt{2}} \big( \textsf{D}_a(\alpha) |1 \rangle_A |\alpha\rangle_B-|\alpha \rangle_A \textsf{D}_b(\alpha)|1 \rangle_B\big).
 \end{equation}
The structure of this state is very simple and follows from displacement of the single-photon entanglement. Nevertheless, it possesses  intriguing properties. It corresponds to a non-Gaussian state which describes the entanglement of two modes, each being individually a mixture of classical and quantum states. The average number of photons $2|\alpha|^2+1$ can be easily adjusted, by varying the amplitude of the initial coherent state. This allows for the exploration of entanglement at various scales ranging from a single photon to macroscopic photon number. Furthermore, the components $\textsf{D}(\alpha) |1 \rangle$ and $|\alpha\rangle$ are easily distinguishable, similarly to the dead and alive components of the famous Schroedinger cat, if one looks at the variance of the photon number: The variance of $\textsf{D}(\alpha) |1 \rangle$ is three times greater than the one characterizing $|\alpha\rangle$ \cite{variance_note}.
This property distinguishes the state (\ref{displ_singlephent}) from another path entangled state \cite{Afek10} obtained by combining a squeezed vacuum $\text{Sq}(\Gamma)\ket{0}$ and a coherent state on a beamsplitter, since the resulting state  $\left(\textsf{D}_a(\alpha)\text{Sq}_a(\frac{\Gamma}{2}) \otimes\textsf{D}_b(\alpha)\text{Sq}_b (\frac{\Gamma}{2})\right)e^{\frac{\Gamma}{2} a^\dag b^\dag}\ket{0,0}$ does not contain components that can easily be distinguished.
The variance of the state (\ref{displ_singlephent}) also makes it a good candidate for precision measurements of phase fluctuations, as we now show.\\

\paragraph{Precision phase measurement} The sensitivity $s$ of a state $\rho_0$ (describing the modes $A$ and $B$) to a phase variation $U_\varphi =e^{i \varphi\, a^\dag a }$ can be estimated through
$
s = ||\frac{\partial \rho_\varphi }{\partial \varphi}||_F^2
$
where $\rho_\varphi= U_\varphi\, \rho_0\, U_\varphi^\dag$ and $||A||_F^2 = \text{tr} A^\dag A.$ One easily shows that $s= - \text{tr} [\,a^\dag a,\, \rho_\varphi]^2$ which reduces to twice the variance of the photon number $s = 2 \,\text{Var}(\,a^\dag a)$ for pure states. Since the variance of the state (\ref{displ_singlephent}) is twice as large as a coherent state with the same mean photon number, its sensitivity is twice as large as a classical state. Moreover, in the presence of loss, the sensitivity of the displaced single photon entanglement becomes $s= 2 \eta |\alpha|^2 (1-3 \eta +4 \eta^2)+\text{cte}$ and beats the coherent state sensitivity ($ s_{\text{cl}}= 2 \eta |\alpha|^2$) if the loss $1-\eta$ does not exceed $\frac{1}{4}.$ For comparision, the sensitivity of the N00N state $\frac{1}{\sqrt{2}}(\ket{N,0}+\ket{0,N})$ ideally scales as the square of the photon number $N$, but it falls down below $s_{\text{cl}}$ (for the same photon number $|\alpha|^2=N/2$) as soon as $\eta < \left(\frac{2}{N}\right)^{1/(2 N-1)}.$ This makes the sensitivity of NOON state very hard to use in practice (eg. $1-\eta$ has to be smaller than $2\%$ $(0.3\%)$ for $N = 100 (1000)$ for NOON states to be useful) and highlight the potential of the state (\ref{displ_singlephent}) for precision phase measurement. We now focus on the robustness of entanglement with respect to loss and phase noise.\\

\paragraph{Robustness with respect to transmission loss.}
In general, entanglement is seen to be increasingly fragile to transmission loss as its size increases. Coherent state entanglement $|\alpha\rangle_A|\alpha\rangle_B + |-\alpha\rangle_A|-\alpha\rangle_B$ \cite{Sanders92} provides a good example. If the mode $B$ is subject to loss (modeled by a beam-splitter with transmission coefficient $\eta_t$) the amount of entanglement, measured by the negativity (see~\cite{negativity, Horodecki09} for definition), decreases exponentially $\mathcal{N} = \frac{1}{2}e^{-2(1-\eta_t)|\alpha|^2}$ with the size $|\alpha|^2$ and loss $1-\eta_t$ ~\cite{coherent_ent}.
In comparison, the state~(\ref{displ_singlephent}) exhibits a surprising robustness. Under the assumption that the mode B undergoes loss, $|\psi_{\rm{out}}\rangle$ becomes a statistical mixture of $\frac{1}{\sqrt{1+\eta_t}} \textsf{D}_a(\alpha) \textsf{D}_b(\sqrt{\eta_t} \alpha) \big(|1 \rangle_A |0\rangle_B-\sqrt{\eta_t}|0 \rangle_A |1 \rangle_B \big)$ and $\textsf{D}_a(\alpha) \textsf{D}_b(\sqrt{\eta_t} \alpha) |0 \rangle_A |0\rangle_B$ with weights $\frac{1+\eta_t}{2}$ and $\frac{1-\eta_t}{2}$, respectively. After applying local displacements $\textsf{D}_a(-\alpha) $ and $\textsf{D}_b(-\sqrt{\eta_t} \alpha)$ to modes $A$ and $B,$ one finds that the negativity of the resulting state is given by $\mathcal{N} = \frac{\eta_t}{2}.$ Since the entanglement cannot increase through local operations, this provides a lower bound for the entanglement before the displacements, i.e. between the macroscopically populated modes. Therefore, the amount of entanglement in $|\psi_{\rm{out}}\rangle$ decays (at worst) linearly with loss, independently of its size. This robustness may be understood in the light of the intimate link between non-classicality and entanglement at a beam-splitter, as mentioned before. Indeed, loss, modeled by a beam-splitter, can be seen as an interaction process entangling the non-classical states and the environment. However, the displacement is a classical operation that does not promote the entanglement of a given quantum system with its environment when it is amplified $(\textsf{D}_b(\alpha)\to \textsf{D}_b(\sqrt{\eta}\alpha)\textsf{D}_{E}(\sqrt{1-\eta}\alpha))$. The robustness of the state~(\ref{displ_singlephent}) makes it well suited for storage in atomic medium. Entanglement between two ensembles containing each a macroscopic number of atoms, has been successfully created by mapping a single-photon entanglement into two atomic ensembles~\cite{ Choi08, Usmani12}. The storage of the displaced single-photon entanglement $|\psi_{\rm{out}}\rangle$ would lead to a similar entanglement in terms of the number of ebits \cite{note_storage}, but it would contain a macroscopic number of excited atoms.\\

\paragraph{Robustness with respect to coupling inefficiency.}
The starting point in our scheme is the creation of a single photon. It is thus natural to ask how the resulting macro entanglement degrades when the single photon is subject to loss (i.e. loss before the 50:50 beam-splitter). For comparison consider for example, micro-macro entanglement obtained by amplifying one photon of an entangled pair \cite{ampli_sti} with an optimal universal cloner~\cite{Lamas-Linares02}. Such entanglement can be revealed even if the amplification is followed by arbitrarily large loss~\cite{Sekatski 2010}. Nonetheless, the state becomes separable as soon as the overall coupling efficiency $\eta_c$ before the cloner is lower than $\frac{n}{n+1},$ $n$ being the average number of photons in the macro component \cite{Sekatski 2010}. On the other hand, one can show following the lines presented in the previous paragraph, that  the negativity of the displaced single-photon entanglement scales like $\mathcal{N}= - \frac{1}{2}\big(1-\eta_c-\sqrt{1-2(1- \eta_c)\eta_c}\big) \approx \frac{\eta_c^2}{4}+\mathcal{O}(\eta_c^3)\geq 0$ where $\eta_c$ stands for the coupling efficiency of the input single photon. This robustness confers to the proposed scheme a great practical advantage over the one based on the universal cloner.\\

\paragraph{Robustness with respect to phase instabilities.}
Another decoherence process for path entanglement is associated with the relative phase fluctuations, due to e.g. vibrations and thermal fluctuations. If the two optical paths corresponding to $A$ and $B$ acquire a phase difference $\varphi,$ the displaced single-photon entanglement becomes
$
|\psi_{\rm{out}}^\varphi\rangle=\frac{1}{\sqrt{2}} \big( \textsf{D}_a(\alpha e^{i\varphi}) e^{i\varphi}|1 \rangle_A |\alpha\rangle_B-|e^{i\varphi} \alpha \rangle_A \textsf{D}_b(\alpha)|1 \rangle_B\big).
$
Furthermore, if $\varphi$ varies from trial to trial, the state  $|\psi_{\rm{out}}^\varphi\rangle \langle \psi_{\rm{out}}^\varphi|$ has to be averaged over $\varphi$ with the probability distribution $p(\varphi)$ associated to the phase noise. The question of the sensibility of the displaced single-photon entanglement with respect to phase instability thus reduces to a measure of the entanglement contained in $\bar{\rho}_{\rm{out}}=\int d \varphi \, p(\varphi) |\psi_{\rm{out}}^\varphi\rangle \langle \psi_{\rm{out}}^\varphi|$. The negativity of this state can easily be obtained numerically by projecting  $\bar{\rho}_{\rm{out}}$ into a finite dimensional Hilbert space (see solid lines in Fig.~2 for a Gaussian probability distribution $p(\varphi)$ with variance $\delta \varphi$). To derive an analytical lower bound on the negativity of this state, we first notice that for any density matrix $\rho$ and any vector $|v\rangle$ the following inequality holds $\langle v| \rho^\Gamma |v\rangle \ge \lambda_{min} \ge -\mathcal{N}_{\rho}$, where $\Gamma$ denotes partial transposition and $\lambda_{min}$ is the smallest eigenvalue of $\rho^\Gamma$. For the state~(\ref{displ_singlephent}), where $p(\varphi)=\delta(0)$,  it is easy to verify that the vector $|v\rangle$ saturating the inequality is $\frac{1}{\sqrt{2}}\textsf{D}_a(\alpha)\textsf{D}_b(\alpha)(|0\rangle |0\rangle-|1\rangle |1\rangle)$. For a general $\bar{\rho}_{\rm{out}}$, it is not optimal, however it provides a lower bound for the estimation of $\mathcal{N}$. We find $\mathcal{N}_{\bar{\rho}_{\rm{out}}}(\delta \varphi,|\alpha|^2) \geq\int d \varphi p(\varphi) \frac{e^{-2 \alpha^2( 2-\cos \varphi)}}{2}  \left( \alpha^2(1-\cos(2 \varphi)) -\cos \varphi \right)$ (the dashed lines in Fig.~2 is obtained by performing the integral numerically for a Gaussian probability distribution). For a Gaussian probability distribution $p(\varphi)$ with variance $\delta \varphi$, the lower bound can be approximated by $ \frac{2-\delta \varphi}{4(1+2|\alpha|^2 \delta\varphi)^{\frac{3}{2}}}$ (see dot-dashed lines in Fig.~2)
\cite{interferometric visibility}. Fig.~2 reveals what is expected from a macroscopic quantum state: the larger the size $2|\alpha|^2+1$ of the state is, the more it becomes sensitive to phase noise.\\

\begin{figure}[h]
\includegraphics[width=8cm]{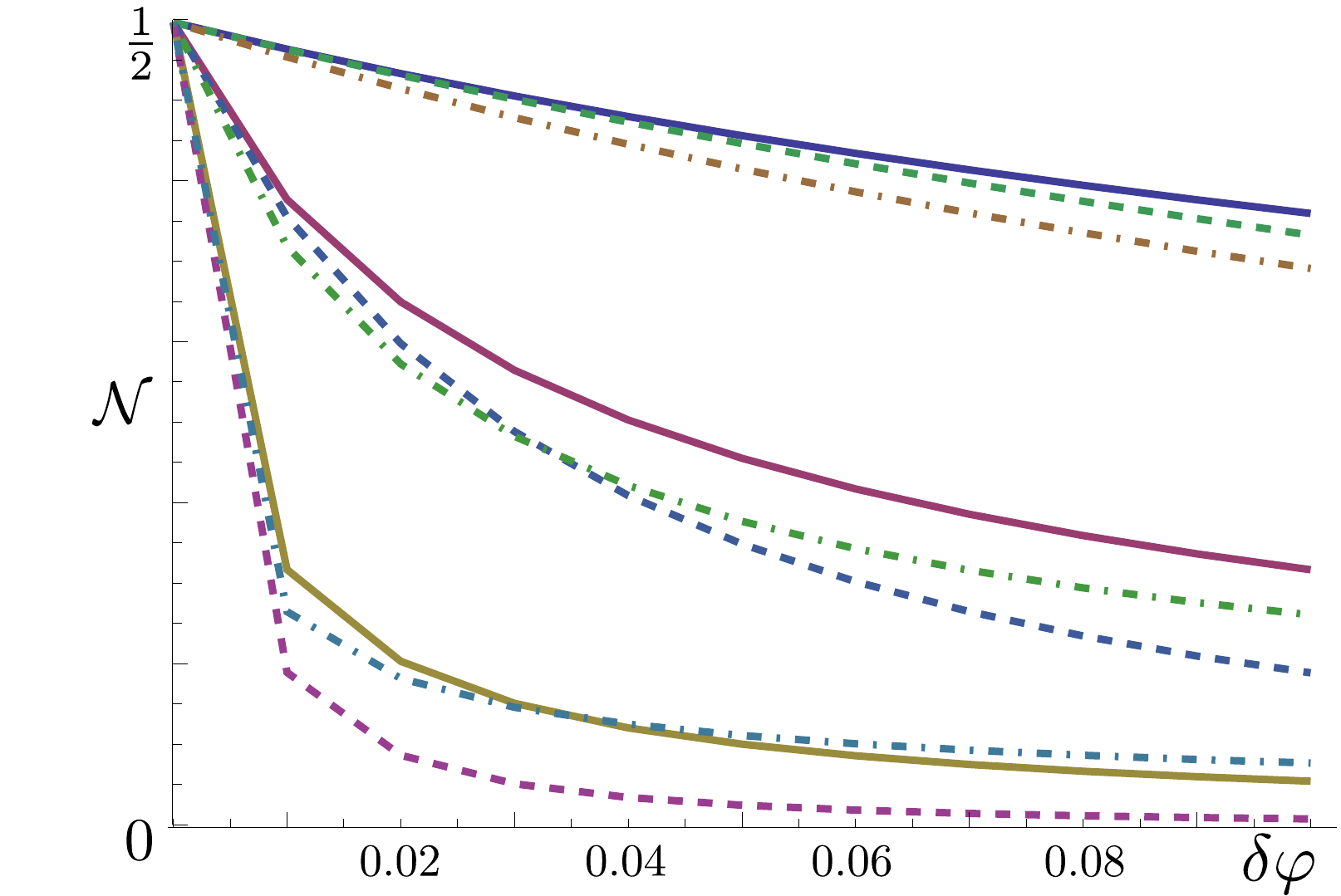}
\caption{Behavior of the negativity of the state $\bar{\rho}_{\rm{out}}$ as the function of the phase noise variance (rad) for various $|\alpha|^2 = \{ 1, 10, 100 \}$ (from top to bottom). The full line is the numerical calculation, the dashed line is a semi-analytical bound, and the dot-dashed line is a fully analytical. (See the main text for details.)}
\label{fig2}
\end{figure}

\paragraph{Revealing displaced single-photon entanglement.} So far, we have discussed the properties of the state (\ref{singlephotonent}). We now present a simple way to reveal its entanglement. The basic idea is to displace each of the electromagnetic fields describing the modes A and B  by $-\alpha.$ Such a displacement in the phase space can be easily performed by mixing the mode to be displaced with an auxiliary strong coherent field (labelled as the local oscillator in the following) on a highly unbalanced beam-splitter~\cite{Paris96}, in a manner similar to homodyne measurements. Since $\textsf{D}_a(-\alpha)=\textsf{D}_a(\alpha)^{-1},$ the modes A and B ideally end up in the state (\ref{singlephotonent}), which can be revealed by tomography using single photon detectors.
Note that a similar approach was proposed in \cite{Raeisi12} to reveal entanglement in the scenario where a macroscopic state is created by phase covariant cloning of an entangled photon pair. The authors proposed to locally undo the cloning before doing the measurement.\\

The approach developed in Ref. \cite{Chou05} does not require a full tomography after the local displacements. It gives a lower bound on the entanglement between the modes A and B from the estimation of the entanglement contained in the two-qubit subspace $\{|00\rangle,|01\rangle, |11\rangle, |10\rangle.\}$. More precisely, the concurrence $C$ of the detected fields is bounded by $C \geq \max\{0, V(p_{01}+p_{10})-2\sqrt{p_{00}p_{11}}\}$~\cite{Chou05}, where $V$ is the visibility of the interference obtained by recombining the modes A and B on a 50:50 beam-splitter, and the coefficients $p_{mn}$ are the probabilities of detecting $m$ photons in A and $n$ in B. This tomographic approach for characterizing single-photon entanglement is attractive in practice and already triggered highly successful experiments, demonstrating e.g. heralded entanglement between atomic ensembles \cite{Chou05, Laurat07, Choi08, Usmani12, Lee11}. \\

Note that the statistical fluctuations in the phase of local oscillators that are used to perform the displacements, limit the precision of the measurement process. Under the assumption that the two local displacements $\textsf{D}_a(-\alpha), \textsf{D}_b(-\alpha)$ are performed with a common local oscillator, the measured state is of the form $\int d\bar\varphi \bar p(\bar\varphi)\textsf{D}_a(-\alpha e^{i\bar \varphi}) \textsf{D}_b(-\alpha e^{i\bar \varphi}) \rho_{\rm{out}} \textsf{D}_a^{\dag} (-\alpha e^{i\bar \varphi})\textsf{D}_b^\dag(-\alpha e^{i\bar \varphi})$ where $\bar p(\bar \varphi)$ stands for the phase noise distribution and $\rho_{\rm{out}}=|\psi_{\rm{out}}\rangle\langle \psi_{\rm{out}}|$. Let $\bar V$ be the visibility of the interference that characterizes the phase stability of the local oscillator. One can show that for small imperfections $\epsilon=(1-\bar V) \ll 1,$ the concurrence is bounded by $ C \geq \max\{0,1-10 (1-\bar V)|\alpha|^2\}.$ The necessary precision of the measurement thus scales as $\frac{1}{\epsilon}=\frac{1}{10 |\alpha|^2}.$ This result strengthens the idea that precise measurements are generally essential for revealing quantum properties of macro systems. \\

\paragraph{Proposed experiment.} We now address the question of the experimental feasibility in detail. For concreteness, we focus on a realization of the single photon source from a pair source based on spontaneous parametric down conversion, the detection of one photon heralding the production of its twin. Filtering techniques must be used so that the mode of the heralded photon can be made indistinguishable from the one of the coherent state produced. (In the supplemental material, we show how one can take mode mismatches into account.) Let $\eta_c$ be the coupling efficiency of the single photon, $\eta_t$ be the global detection efficiency, including the transmission from the 50:50 beam-splitter to the detector, as before. For small heralding efficiency and if the parametric process is weakly pumped so that the success probability for the emission of one photon pair is small, one can show that the concurrence is bounded by $C\geq \max\{0,\eta - 2\sqrt{2\eta(1-\eta)} \sqrt{\epsilon \eta_t|\alpha|^2}-2(2+3\eta)\epsilon \eta_t|\alpha|^2\}.$ Here $\eta=\eta_c\eta_t$ and $\epsilon=1-V$ where $V$ is the interferometric visibility that characterizes the phase stability of $A$, $B$ and the local oscillator (we assume that the modes have independent phase fluctuations with the same variance). To know the value of the visibility that can be obtained in practice, we built a balanced Mach-Zehnder interferometer (See supplemental material). Using an active stabilization, we measured a visibility of $V=99.996\pm 0.001$\%. Assuming a coupling efficiency $\eta_c = 50$\% and a detection efficiency $\eta_t = 60$\%, the concurrence remains positive $(C\approx 0.01)$ for $|\alpha|=28.$ This translates into entanglement populated by more than $(2|\alpha|^2+1)=1500$ photons.\\

\paragraph{Conclusion.}
We have presented a scheme for creating and revealing macroscopic entanglement with a single photon, coherent states and linear optical elements. The simplicity of our proposal is conceptually remarkable. On the one hand, it highlights the idea that although quantum systems are difficult to maintain and observe at macro scales, they can easily be created. On the other hand, it naturally raises the question: is the resulting state really macroscopic? We have shown through experimental results that the entangled state that could be obtained with currently available technologies would involve a large enough number of photons to be seen with the naked eyes~\cite{SekatskiPRL}. This makes our approach satisfactory if macroscopicness is a notion related to the size. We also mentioned that the components of the entangled state can easily be distinguished with a mere avalanche photodiode if one looks at the variance of the photon number distribution. This pleases the ones who believe that macro entangled states need to have components that can be easily distinguished. Although our study showed that the resulting state features an unexpected robustness against loss, we have shown that it is also more and more fragile under phase disturbance when its size increases. Our approach is thus also satisfactory if macroscopic means sensitive to decoherence and highlights the complexity of possible interactions between a given quantum system and its surroundings. We have also seen that the precision of the measurement that is required to reveal the quantum nature of the produced state increases with its size. This also makes our scheme satisfactory if macroscopicness is related to the requirement on the measurement precision. Note to finish that there are many other candidates for macroscopicity measure \cite{Frowis12}. Testing each of them is work for future.\\

We thank C. Fabre, J. Laurat, A.I. Lvovsky, A. Martin, M. Mitchell, O. Morin, H. de Riedmatten, C. Simon, R. Thew, H. Zbinden for helpful comments, and we acknowledge the Swiss NFS and the EU project Qessence for financial support. M.S. was supported supported by the EU 7FP Marie Curie Career Integration Grant No. 322150 "QCAT" and by MNiSW grant No. 2012/04/M/ST2/00789.\\


\begin{thebibliography}{3}
\bibitem{note_ampli} In the presence of loss before and after the phase covariant cloner, we do not know when the amplified state becomes separable.
\bibitem{negativity} Negativity $\mathcal{N}$~\cite{Horodecki09} conveniently quantifies entanglement. It is defined by $\mathcal{N} = \sum_{\lambda<0} |\lambda|$,  where $\lambda$ are the eigenvalues of the partially transposed density matrix. Accordingly to this definition the negativity of a two qubits states if bounded by $1/2$.
\bibitem{coherent_ent} Consider transmission efficiency $\eta$ such that $\eta |\alpha|^2 \gg 1.$ Further consider that the mode B undergoes loss. The coherent state entanglement becomes $|\alpha\rangle_a|\eta \alpha\rangle_b|\sqrt{1-\eta} \alpha\rangle_{\ell}+|-\alpha\rangle_a|-\eta \alpha\rangle_b|-\sqrt{1-\eta} \alpha\rangle_{\ell}$ where $\ell$ stands for the lost mode. Tracing $\ell$ out, one gets
\begin{eqnarray}
\nonumber
&\frac{1}{2}\Big(&|\alpha\rangle_a|\sqrt{\eta}\alpha\rangle_b \langle \alpha|_a\langle \sqrt{\eta}\alpha|_b| \\
\nonumber
&&+|-\alpha\rangle_a|-\sqrt{\eta}\alpha\rangle_b \langle -\alpha|_a\langle -\sqrt{\eta}\alpha|_b\\
\nonumber
&&-e^{-2(1-\eta)|\alpha|^2}|\alpha\rangle_a|\sqrt{\eta}\alpha\rangle_b\langle -\alpha|_a\langle -\sqrt{\eta}\alpha|_b\\
\nonumber
&&
-e^{-2(1-\eta)|\alpha|^2}|-\alpha\rangle_a|-\sqrt{\eta}\alpha\rangle_b\langle \alpha|_a\langle \sqrt{\eta}\alpha|_b \Big).
\end{eqnarray}
The smallest eigenvalue associated to its partial transpose is $-e^{-2(1-\eta)|\alpha|^2}.$
\bibitem{note_storage} Note that if multimode memories are used, the number of ebits could easily be increased by combining time-bin qudits with a bright coherent state on a beam-splitter.
\bibitem{interferometric visibility} In practice, the phase stability is usually characterized by the classical interferometric visibility $\bar V$  that would be obtained by recombining the mode A and B on a 50:50 beam-splitter~\cite{Minar}. Using the relation between the visibility and the phase fluctuation  $ V=e^{\frac{-\delta \varphi}{2}} \approx 1-\frac{\delta \varphi}{2}$, we obtain $\mathcal{N}_{\bar{\rho}_{\rm{out}}}( V, |\alpha|^2) \geq \frac{ V}{2(1+4|\alpha|^2(1-V))^{\frac{3}{2}}}$.
\end{thebibliography}
\end{document}